\documentclass[sigconf,nonacm]{acmart}

\usepackage{amsmath,amsfonts}
\usepackage{algorithm}
\usepackage{algorithmic}
\usepackage{graphicx}
\usepackage{multirow}
\usepackage{booktabs}
\usepackage{xspace}
\usepackage{enumitem}
\usepackage{pifont}
\usepackage{subcaption}

\usepackage{algorithmic,algorithm}
\usepackage{booktabs}
\usepackage{multirow}
\usepackage{graphicx}
\usepackage{textcomp}
\usepackage{xcolor}
\usepackage{enumitem}
\usepackage{siunitx}
\usepackage{pifont}
\usepackage{ifthen}
\usepackage{subcaption}

\newcommand{\bh}[1]{\vspace{3pt}\noindent \textbf{#1} }
\newcommand{\ih}[1]{\vspace{1pt}\noindent \textit{#1} }

\usepackage{xcolor}
\usepackage{soul}
\usepackage{ifthen}
\usepackage{lipsum}

\definecolor{newtext}{rgb}{0.1, 0.1, 0.9} 
\newboolean{reviewmode}
\setboolean{reviewmode}{true} 

\newcommand{\newtext}[1]{%
    \ifthenelse{\boolean{reviewmode}}{%
        \textcolor{newtext}{#1}%
    }{%
        #1%
    }%
}
\newcommand{\deletedtext}[1]{%
    \ifthenelse{\boolean{reviewmode}}{%
        \st{#1}%
    }{%
    }%
}
\def\BibTeX{{\rm B\kern-.05em{\sc i\kern-.025em b}\kern-.08em
    T\kern-.1667em\lower.7ex\hbox{E}\kern-.125emX}}

\begin{document}


\title{\textbf{LEXI}: \textbf{L}ossless \textbf{Ex}ponent Coding for Efficient \textbf{I}nter-Chiplet Communication in Hybrid LLMs}


\author{Miao Sun, Alish Kanani, Kaushik Shroff, and Umit Ogras}
\affiliation{%
  \institution{Department of Electrical and Computer Engineering, University of Wisconsin-Madison}
  \city{}
  \state{}
  \country{}
}
\email{{smiao23, ahkanani, kshroff, uogras}@wisc.edu}

\renewcommand{\shortauthors}{Sun et al.}


\begin{abstract}
Data movement overheads increase the inference latency of state-of-the-art large language models (LLMs). These models commonly use bfloat16 (BF16) format for stable training. 
Floating-point standards allocate eight bits to the exponent, but our profiling reveals that exponent streams exhibit fewer than 3 bits Shannon entropy, indicating high inherent compressibility. To exploit this potential, we propose LEXI, a novel \textit{lossless exponent compression scheme} based on Huffman coding. 
LEXI compresses activations and caches on the fly while storing compressed weights for just-in-time decompression near compute, without sacrificing system throughput and model accuracy. The codecs at the ingress and egress ports of network-on-chip routers sustain the maximum link bandwidth via multi-lane LUT decoders, incurring only 0.09\% area and energy overheads with GF 22 nm technology. 
LEXI reduces inter-chiplet communication and end-to-end inference latencies by 33–45\% and 30–35\% on modern Jamba~\cite{lieber2024jamba}, Zamba~\cite{glorioso2024zamba}, and Qwen~\cite{bai2023qwen} LLMs implemented on a homogeneous chiplet architecture. 
\end{abstract}

\maketitle
\vspace{-3mm}
\section{Introduction} \label{sec:introduction}


Increasing demand for LLM-based generative pre-trained transformer (GPT)-class services pressures compute and storage resources. 
For example, Llama~\cite{touvron2023llama}, Falcon~\cite{almazrouei2023falcon}, and Qwen~\cite{bai2023qwen} models have scaled to hundreds of billions of parameters, while their computational loads have grown from approximately 100 TeraFLOPS to 1 PetaFLOP per inference. 
The inference latency is increasingly dominated by data movement as (i) the sheer size of model parameters stresses memory capacity and bandwidth, and (ii) the key–value (KV) cache inherent to self-attention grows with sequence length~\cite{dao2022flashattention,kwon2023efficient}. 
These challenges are amplified in multi-die/chiplet accelerators, where cross-die traffic during the decode phase becomes a first-order bottleneck~\cite{sharma2025heterogeneous}.
Quantization and pruning can remedy these problems, but they sacrifice model accuracy for efficiency~\cite{tao2019dps,han2016deep}.

Hybrid models that integrate Transformer and Mamba modules have recently emerged as promising solutions toward balancing memory footprint, computational intensity, performance, and accuracy~\cite{lieber2024jamba, glorioso2024zamba, blakeman2025nemotron}. 
These architectures reduce KV cache pressure by substituting some attention blocks with fixed-size (sequence-length independent) state–space layers~\cite{gu_Mamba_2023}.
Despite these improvements, inter-die communication overheads remain a first-order bottleneck~\cite{xu2025wsc}. 
Hence, this paper targets the data transfer overhead, a critical issue because LLM execution is fundamentally limited by memory bandwidth in data-intensive phases~\cite{sun2023length, moitra2025meadow}.

\begin{figure*}[t]
\centering
\includegraphics[width=0.85\linewidth]{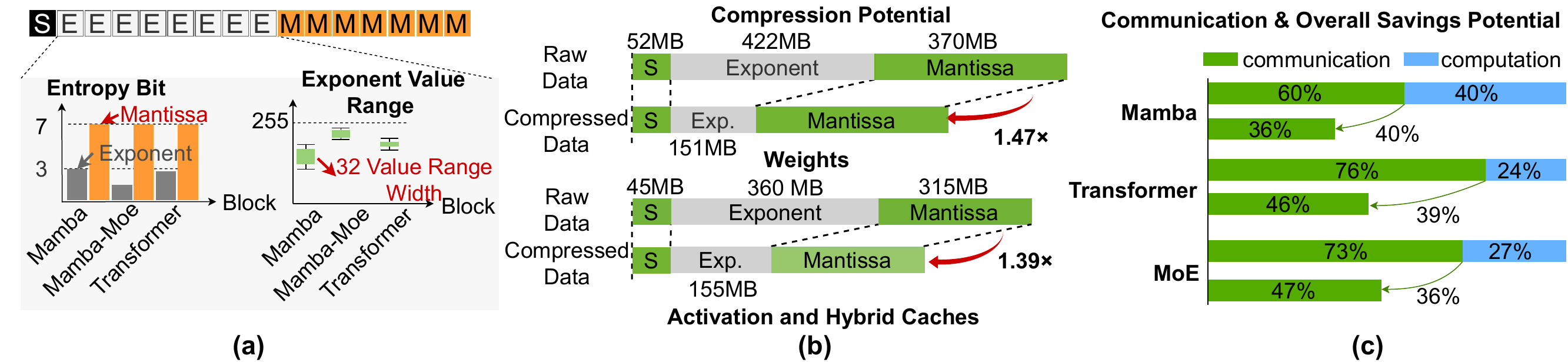}
\vspace{-4mm}
\caption{Exponent compression opportunities on NVIDIA GeForce RTX 3090~\cite{NVIDIA_RTX3090}.
(a) BF16 exponents show only $\sim$3 bits of Shannon entropy and span fewer than 32 values, while mantissas consistently use the full 7-bit range.
(b) Exponent compression shrinks weights from 422 MB to 151 MB and activations/caches from 360 MB to 155 MB.
(c) Lossless exponent compression reduces communication overhead by 36-40\% across Mamba, Transformer, and MoE blocks.
}

\label{fig:model-profiling}
\vspace{-4mm}
\end{figure*}



The BF16 format has become the dominant precision for neural network training and inference, particularly in GPU- and cloud-computing environments. 
For example, the Hugging Face Model Hub, an industry standard platform, declared BF16 as the default precision for most widely adopted LLMs. Indeed, the configuration metadata of prominent LLMs, such as the Llama and Mistral series, specify the \texttt{torch\_dtype} as \texttt{"bfloat16"}. 
Using 8-bit exponents, BF16 achieves the same dynamic range as FP32, mitigating overflow and underflow without the explicit loss scaling often required by FP16. 
However, our profiling shows that the exponent field exhibits on average fewer than 3 bits Shannon entropy during inference due to the limited span of the weights and runtime activations. \textit{This phenomenon offers an opportunity to reduce communication costs without altering numerical accuracy.}

To exploit this potential, we propose LEXI, a lossless exponent coding for efficient inter-chiplet communication. 
LEXI compresses exponents while preserving the numerical representation of BF16 in memory and compute units, as illustrated in Figure~\ref{fig:model-profiling}. 
It transfers the weights and activations in a compact lossless form, thereby reducing data-movement latency across memory hierarchies and chiplet interconnects. 
The design has two paths: (i) offline compression of weights stored in DRAM/high-bandwidth memory (HBM), and 
(ii) on-the-fly compression of activations, KV caches in attention blocks and state caches in SSM blocks, collectively referred to as \textit{hybrid caches}. 
A novel lightweight multi-cache Huffman encoder maintains exponent frequency statistics, generates per-layer codebooks, and compresses exponent streams at runtime, with a small codebook piggybacked alongside the bitstream. 
On the receiving side, a multi-stage lookup table-based decompression provides a single-cycle decode, maintaining pipeline throughput. 

We implemented LEXI using GlobalFoundries (GF) 22 \si{\nano\meter} technology and integrated it 
on a representative homogeneous multi-chiplet architecture~\cite{shao2019simba}. 
Since LEXI affects only inter-chiplet and memory-chiplet communication, it is orthogonal to the computations within the chiplet. In our setup, we simulate communication traffic generated by three recent and representative LLMs: Jamba-tiny-dev~\cite{lieber2024jamba}, Zamba2-1.2B-Instruct-v2~\cite{glorioso2024zamba}, and Qwen1.5-1.8B-Chat~\cite{bai2023qwen} across WikiText-2~\cite{merity2016wiki} and Colossal Clean Crawled Corpus (C4)~\cite{raffel2020c4} datasets. LEXI reduces communication latency by 33–45\% and end-to-end inference latency by 30–35\%, all while maintaining lossless operation.
Our GF 22 \si{\nano\meter} implementation shows only 0.09\% area overhead for these runtime compression /decompression units, highlighting LEXI's practicality.

To our knowledge, this work is the first to show that exponent-only lossless coding can be implemented at full speed with negligible area and energy overhead.
Our major contributions are:
\begin{itemize}[leftmargin=*]
\item A novel exponent-only compression scheme for BF16 format, which targets the weights, activations, and hybrid caches in state-of-the-art LLMs.
\item Efficient hardware implementation of exponent-only codecs and their seamless integration at the NoC routers and I/O with negligible area overhead using GF 22 \si{\nano\meter} technology.
\item Comprehensive performance evaluations on a multi-chiplet platform, demonstrating significant performance gains and up to 35\% lower end-to-end inference latency.
\end{itemize}


\vspace{-1mm}
\section{Related Work} \label{sec:Related work}

The growing scale of LLMs has shifted inference bottlenecks from compute to data movement, especially during decode phase~\cite{recasens2025mind,afifi2024accelerating}.
Generating a single token in this phase requires accessing all model weights and hybrid caches from the prior context, creating a disparity between processor speed and available memory bandwidth, commonly referred to as the ``memory wall''~\cite{gholami2024ai,mutlu2022modern}. 
This phenomenon reflects the low compute intensity of LLM workloads, where the ratio of FLOPs to bytes transferred is insufficient to hide memory latency, resulting in the under-utilization of compute resources.

Hybrid LLM models complement Transformer attention layers with Mamba SSM layers with fixed-size recurrent state, alleviating memory requirements~\cite{gu_Mamba_2023,dao2024transformers}. 
For example, Jamba~\cite{lieber2024jamba} and Zamba~\cite{glorioso2024zamba} achieve 3$\times$ and 1.6$\times$ faster inference than Mixtral~\cite{jiang2024mixtral}.
On edge computing platforms, as demonstrated by eMamba~\cite{kim2025emamba}, which achieves $2.22\times$ and $9.95\times$ higher throughput compared with Transformer and CNN, respectively.
However, algorithmic improvements alone are insufficient to minimize the communication overheads since hybrid models remain memory-bound during inference due to fundamentally autoregressive decode phases. 
Quantization of weight and activations all the way to extreme two bits~\cite{xue2024oltron, liu2025paretoq} and pruning~\cite{huang2024new} have commonly been used to reduce model size, area and communication overhead. However these approaches sacrifice the model performance. In contrast, LEXI addresses this gap through a hardware-implemented, lossless compression scheme that targets exclusively the exponent field of BF16 data. As an orthogonal approach, LEXI can be seamlessly combined with existing quantization and pruning techniques. Hence, it can be applied after any quantization or pruning technique to further reduce the communication overhead \textit{without impacting the model performance}.

\begin{table}[!b]
    \footnotesize 
    \centering
    \vspace{-4mm}
    \caption{Methodology comparison against related work.}
    \vspace{-4mm}
    \label{tab:comparison} 
    \begin{tabular}{l @{\hspace{3pt}} c @{\hspace{6pt}} p{3.0cm} @{\hspace{6pt}} c}
        \toprule 
        \textbf{\raggedright Work} & \textbf{Lossless} & \multicolumn{1}{c}{\textbf{Compressed Data}} & \textbf{Implementation} \\
        \midrule 
        \raggedright HACK~\cite{zhang2025hack} & \ding{55} & \centering KV-cache & SW \\
        \raggedright KVComp~\cite{jiang2025kvcomp} & \ding{55} & \centering KV-cache & SW \\
        \raggedright Ecco~\cite{cheng2025ecco} & \ding{55} & \centering KV-Cache/Act./Weight & HW \\ 
        \raggedright Zhang et. al~\cite{zhang202570} & \ding{51} & \centering Weight & SW \\
        \raggedright Huff-llm~\cite{yubeaton2025huff} & \ding{51} & \centering Weight & SW \\
        \raggedright Zipnn~\cite{hershcovitch2025zipnn} & \ding{51} & \centering Weight & SW \\
        \midrule 
        \textbf{\raggedright LEXI} & \textbf{\ding{51}} &  \multicolumn{1}{c}{\textbf{KV-Cache/Act./State/Weight}}  & \textbf{HW} \\
        \bottomrule
    \end{tabular}
     \vspace{-2mm}
\end{table}

Data compression has been effective in reducing on-chip network traffic~\cite{raparti2018dapper,wang2016disco}.
Several lossy and lossless compression techniques have been proposed to reduce the storage and communication costs of deep learning models.
As shown in Table~\ref{tab:comparison}, prior work has largely focused on lossy compression~\cite{zhang2025hack,jiang2025kvcomp, cheng2025ecco}, which degrades the performance of the original models. Moreover, existing work on lossless compression~\cite{zhang202570, yubeaton2025huff,hershcovitch2025zipnn} has only considered weight compression (e.g., for loading model parameters from the HBM) without addressing the cross-die communication cost. 
Furthermore, software (SW) compression schemes do not address the dataflow and interconnect bottlenecks of chiplet-based accelerators, where inter-chiplet communication remains a primary constraint. Thus, the focus shifts to dedicated hardware (HW) solutions.

In contrast to prior work, LEXI applies \textit{lossless compression to weights, activations, and hybrid caches}. We present a low-cost hardware implementation to the communication latency while avoiding compute overhead. 
Our novel design requires a small alphabet and lookup tables (LUTs) with only 0.09\% area overhead while achieving full speed (one exponent/cycle) without any performance loss.


\vspace{-2.5mm}
\section{Exponent Statistics Profiling and Motivation} \label{sec:III}

We first analyze the bit-level entropy in weights, activations, and hybrid caches
to analyze potential savings and guide the design process.
This profiling exposes redundancies in BF16 representation during inference and highlights the key challenges for practical, hardware-efficient compression.

\vspace{-2.5mm}
\subsection{\hspace{-3mm}Profiling Methodology and Illustrative Results}
We run the target workloads on a workstation with an NVIDIA GeForce RTX 3090 GPU~\cite{NVIDIA_RTX3090}. 
Our instrumented inference script logs (i) weights, (ii) activations, and (iii) hybrid caches (i.e., the Transformer KV cache and the Mamba state cache) at layer boundaries. 
At execution, each BF16 value is parsed into its \{\texttt{sign, exponent, mantissa}\} fields and computes the entropy for the exponent stream. 

Profiling the Jamba-tiny-dev model over a 1K-token WikiText-2 sequence (Fig.~\ref{fig:model-profiling}) shows that the BF16 exponent stream exhibits at most $\sim$3 bits of entropy, indicating substantial lossless compressibility.
This stems from their narrow dynamic range, with most exponent values \textit{concentrated within 32 distinct values.}
In contrast, the mantissa consistently uses its full 7-bit precision and offers little compression potential.

Leveraging this property, as shown in Fig.~\ref{fig:model-profiling}(b), LEXI can reduce the exponent volume from 422 to 151 MB for weights and 360 to 155 MB for activations and hybrid caches, corresponding to overall data-volume reductions of 1.47$\times$ and 1.39$\times$, respectively.
As presented in Fig.~\ref{fig:model-profiling}(c), these savings translate directly into communication cost reduction by 40\%, 39\% and 36\% in Mamba, Transformer, and mixture-of-experts (MoE) blocks, respectively.

\subsection{Motivation and Design Challenges}
The profiling results indicate that exploiting the redundancy in BF16 \emph{exponents} can substantially decrease the communication overhead. Realizing this potential during inference requires tackling two primary challenges.

\bh{Challenge 1 - Real-time data compression:} 
Compression and decompression must sustain link-rate throughput with a short deterministic latency so that their overheads do not outweigh the reduced transfer time.
This requires a streaming operation with (1) flit-aligned framing, (2) fast dynamic codebook configuration, and (3) decoder latency compatible with ingress timing. This demands transforming complex Variable-Length Coding into a predictable cycle-count hardware pipeline. The compression scheme employs a dedicated circuit with minimal internal dependencies to ensure true link-rate throughput, mitigating the scheduling challenges caused by variable output size.

\bh{Challenge 2 - Hardware overhead:} 
The compression and decompression units must incur minimal area overhead, avoiding wide datapath overheads and large buffer costs for a chiplet-ready design. Conventional entropy coders are impractical due to sizable dictionaries/SRAMs, deep FIFOs, and long critical paths \cite{ledwon2019design,liu2010architecture}, which inflate area and dynamic power. A chiplet-ready design must avoid wide datapath overheads and meet frequency targets without large buffer costs. This requires minimal-state, context-aware encoding optimized for BF16 exponent content, favoring minimal-sized dictionaries. Achieving high clock speed relies on shallow, heavily pipelined logic to maintain a short critical path, structuring the final hardware to rely primarily on combinational logic rather than large storage structures.

The proposed LEXI framework is designed to solve these challenges. 
Its streaming, exponent-aware, and hardware-efficient design exploits bit-level redundancy without altering BF16 semantics.



\vspace{-2mm}
\section{LEXI: Entropy-Aware Exponent Compression} \label{sec:IV}
\vspace{-1mm}





\subsection{\hspace{-2mm}Definitions and LEXI Architecture Overview}\label{ssec:overview}
\vspace{-1mm}

\begin{figure}[t]
\centering
\includegraphics[width=1\linewidth]{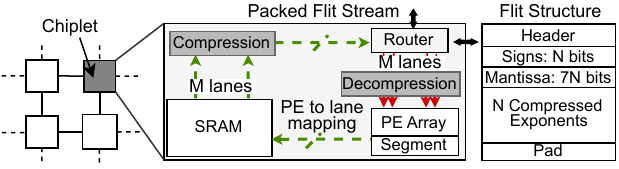}
\vspace{-7mm}
\caption{Overview for proposed compression/decompression units in chiplet system.}
\vspace{-5mm}
\label{fig:overview}
\end{figure}

\bh{Interfacing with Processing Elements (PEs) within chiplets:} 
The proposed LEXI framework is oblivious to the computations within each PE. 
The compression and decompression circuits reside on the interconnection network egress and ingress paths as shown in Figure~\ref{fig:overview}.
\textbf{Activations and hybrid caches} are produced at runtime at compute chiplets in standard (decompressed) format. 
To minimize the transfer time to their destinations, the exponent components are compressed on-the-fly at egress, before transmission. Then, the receiving unit (compute chiplet) decompresses them before using.
In contrast, \textbf{the model weights} are assumed to be compressed offline and stored in compressed format in memory to reduce their footprint. When requested, the memory unit transmits them in this format. The receiving compute chiplet decompresses then at ingress on-the-fly before using.

\bh{Huffman tree generation boundary:}
Generating a new Huffman tree for each layer's output increases locality (hence, compression ratio) \textit{as opposed to using a single tree across the entire inference execution}. 
Once a PE array in a chiplet finishes computations, it produces a fixed number of activations per layer (e.g., 2M activations on average for Qwen1.5-1.8B-Chat model with 1K input tokens when it is mapped to our representative homogeneous chiplet architecture~\cite{shao2019simba} in our experimental evaluations). We initiate tree generation with the first 512 activations, incurring the $78-cycle$ penalty only once. The subsequent $\sim$1.9M of activations leverage this tree with zero overhead. 
Empirical results show that the exponent distribution remains stable across a layer’s outputs because repeated layer-normalization operations keep the activations within a bounded and consistent range.
In this way, the proposed LEXI framework (1) benefits from a custom Huffman tree for each set of layer outputs, and (2) incurs a small one-time cost while leveraging the same tree for all remaining activations.

\bh{Flow control unit (flit):}
On-chip and inter-chiplet networks transmit packetized data in packets in units called flits (one flit is sent per cycle). 
To sustain this maximum link bandwidth, LEXI puts the compressed activations into flits with the following format: 
$$\{\texttt{Header, Sign bits, Mantissa, Compressed Exponents}\}$$

The Header indicates how many activations are sent in a given flit, as shown in Figure~\ref{fig:overview}. If the exponent stream does not end on a flit boundary, the remainder is zero-padded. Our design enables streaming flit-aligned transfers without intermediate expansion.
At the ingress, the decoder reads the header and reassembles the decompressed activations before passing them back to the PEs.

\begin{figure*}[t]
\centering
\includegraphics[width=1\linewidth]{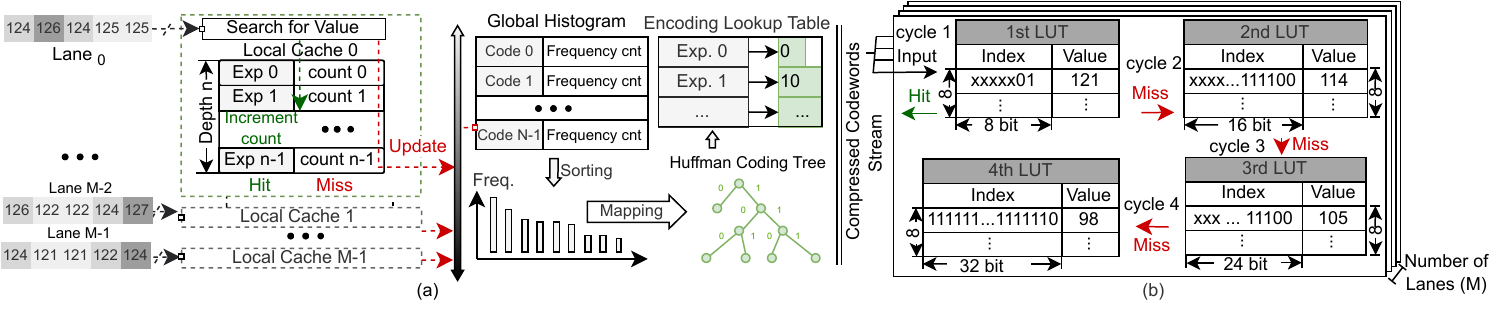}
\vspace{-7mm}
\caption{Hardware microarchitecture of LEXI: (a) compression circuit with $M$ parallel lanes, where per-lane local caches accelerate Huffman tree creation and a simple lookup table encodes exponents; (b) Four-stage decompression unit, each stage containing eight entries indexed by 8-, 16-, 24- and 32-bit indices.}
\vspace{-3mm}
\label{fig:encoder}
\end{figure*}

\vspace{-2mm}
\subsection{Runtime Huffman Tree Generation}\label{ssec:compression}

LEXI generates a Huffman tree in two steps: 1) constructing a global histogram of exponent frequencies, and 2) codebook generation.

\subsubsection{Parallel Histogram Generation:}
LEXI compresses the activations on-the-fly at the egress boundaries as PEs send them to the interconnection network, as discussed in Section~\ref{ssec:overview}.
To prevent throughput bottlenecks, as shown in Figure~\ref{fig:encoder}, it distributes the exponent counting process for histogram generation across $M$ parallel lanes, each equipped with its own local cache.
The optimal value of $M$ is found through design space exploration (Section~\ref{ssec:trade-off}).

\bh{Operations at each lane:}  
As the exponents arrive from the PE array, the first step is checking whether they are already in the local cache with a current count.
If the newly arrived exponent is already in the local cache (\textit{a cache hit}), the lane simply increments its local counter. 
Our novel $M$-lane design ensures high locality and hit rate at each lane-specific local cache.
In case of a \textit{cache miss}, the oldest exponent is evicted from the local cache and written to the global histogram with its accumulated count. 
Then, the new entry is written to the cache with a count of one. 
Our design space exploration in Section~\ref{ssec:trade-off} shows that a cache of eight entries per lane optimizes the balance hit rate and area cost.

\bh{Global histogram updates:} 
The global histogram accumulates exponent counts from parallel write operations by the local caches. Competition for histogram port access is resolved via a simple arbiter. The arbiter grants exclusive use to the first arriving request for a fixed duration of three cycles before release. The merged global histogram is then read by the Huffman tree builder to emit the codebook.

\vspace{-2mm}
\subsubsection{Pipelined Codebook Generation Hardware}
After receiving all activations required for the Huffman tree generation, the global histogram contains the frequency counts for all observed exponents. Profiling results show consistently fewer than $\mathbf{32}$ distinct exponent values. Hence, the primary pipeline is designed for this $\mathbf{32}$-entry range; exceptional values are managed via a fallback mechanism (as detailed at the end of this section). With this data, the tree generation proceeds in three pipelined stages:

\bh{1. Sorting the exponents in descending order of their counts:} We employ a parallel bitonic sorter~\cite{batcher1968sorting} optimized for small input sizes ($\leq$32 elements). 
The sorter requires $log_2(32) \times (log_2(32)+1)/2 = 15$ stages, completing in 15 cycles. This balances area efficiency with throughput requirements.

\bh{2. Huffman coding tree construction:} Following canonical Huffman procedures~\cite{4051119}, we build the code tree by repeatedly merging the two least-frequent symbols. Our hardware implementation uses a priority queue backed by the sorted frequency list, requiring 31 cycles for worst-case tree construction (32 elements).

\bh{3. Code assignment and LUT programming:} The tree is traversed to generate prefix-free codes, with code lengths stored in the encoding LUTs across all $M$ lanes. This requires 32 cycles to program all LUT entries. The entire process completes in 78 cycles (15+31+32), which is seamlessly pipelined with the incoming data stream for on-the-fly compression.

\bh{Exception handling for guaranteed functional correctness:} To ensure correctness against raw exponents exceeding the $\mathbf{32}$-entry encoding limit, a lookup miss triggers a $\mathbf{32}$-bit fallback: a reserved $\mathbf{24}$-bit escape code (all ones) followed by the raw $\mathbf{8}$-bit exponent. Since such exceptions are extremely rare (not observed in our experiments), this mechanism imposes negligible bandwidth overhead while maintaining functional integrity.
\vspace{-2mm}
\subsection{Parallel Stream Encoding and Transmission}\label{sec:parallel_stream}
Following the pipelined codebook generation, LEXI immediately commences continuous, high-throughput encoding.

\bh{Distributed parallel lookup:} To eliminate lookup contention and maximize throughput, LEXI replicates the 32-entry encoding LUT constructed during codebook generation at each parallel lane. 
This ensures all $M$ lanes perform simultaneous, single-cycle lookups, transforming the 8-bit exponent into its \texttt{codeword}. 

\bh{Flit-aligned packetization:} A critical step for sustaining the maximum throughput is aggregating the variable-length \texttt{codewords} efficiently. 
The network interface receives the stream of \texttt{codewords} from all lanes, along with their associated sign and mantissa bits, and packs them into fixed-size flits. Since Huffman codes are \textbf{prefix-free}, a minimal, per-layer codebook header is simply prepended to the compressed stream for lossless decoder reconstruction.

\bh{Sustained line rate operation:} \textit{The entire data path forms a non-blocking pipeline, sustaining one-flit-per-cycle output rate for full inter-layer link bandwidth utilization}. Although an initial startup latency of $78$ cycles ($78~ns$ under $1~GHz$ in our experiments) is incurred once per layer, this cost is negligible against millisecond-scale communication of millions of activations. Crucially, the effective overhead vanishes and maximum link utilization is maintained 
because all encoding steps (histogram accumulation, tree creation, LUT programming) are fully pipelined with subsequent data.

\vspace{-2mm}
\subsection{On-the-fly Decompression Circuit}
Since Huffman decoders process a variable-length bitstream, finding codeword boundaries is inherently sequential. A naive implementation uses a single large LUT indexed by $L_{max}$(24 bits) to return $\{\text{\texttt{exponent, bit-length}}\}$ in one cycle, but incurs prohibitive area cost and poor scalability.
Thus, we design a multi-stage decompression circuit ($\text{Figure~\ref{fig:encoder}(b)}$). This multi-stage LUT segments the codebook based on frequency and code length. Stage 1 indexes the first $B_1$ bits; it returns the symbol if available, else redirects to the next stage.
Each subsequent stage k then consumes the additional $B_k$ bits and repeats until the symbol is resolved.
Since shorter, high-frequency codes are often resolved in Stage 1 (one cycle), only longer, rarer codes traverse deeper stages.

The key design choice lies in determining the optimal bit-width for each stage. 
Through extensive design space exploration across all target workloads, we found that a four-stage configuration with $B_1, B_2, B_3, B_4 = 8,16,24,32$ bits provides the optimal balance between area efficiency and decoding latency, as detailed in Section~\ref{ssec:trade-off}.
The final stage can also program the reserved escape code. 
To sustain required throughput, we instantiate multiple parallel decode lanes, distributing incoming flits in a round-robin manner for independent, parallel decoding. 
This configuration proved area-efficient and line-rate capable, as analyzed in Section~\ref{ssec:trade-off}. 

\section{Implementation and Evaluation of LEXI} \label{sec:V}
\subsection{Experimental Setup}


\ih{Models and datasets:}
The proposed compression technique is evaluated using two hybrid models Jamba-tiny-dev (319M)~\cite{lieber2024jamba}, Zamba2-1.2B-Instruct-v2~\cite{glorioso2024zamba}) and one transformer-only model (Qwen1.5-1.8B-Chat~\cite{bai2023qwen}).
In the remainder of this section, we refer to these models as Jamba, Zamba, and Qwen for simplicity.
All three models are tested using WikiText-2~\cite{merity2016wiki} and C4~\cite{raffel2020c4} datasets, with input sequence lengths of 1K and 2K tokens, respectively. In both cases, the output sequence length is fixed at 512 tokens.

\bh{Target hardware:}
To demonstrate LEXI, we integrated it into a 6$\times$6 homogeneous chiplet array based on the Simba~\cite{shao2019simba} architecture. This test case is selected since it was validated with hardware implementation. 
It uses a 2D mesh network-on-interposer (NoI) with 100 $\text{Gbps}$ inter-chiplet links. 
Each chiplet executes block-level kernels (Mamba/SSM, attention, and feed-forward).
LEXI transmits compressed weights and decompresses the weights on-the-fly before use. Hybrid caches are compressed block-by-block when written back to memory, then retrieved and decompressed just prior to computation. 
Activations are transmitted between chiplets in compressed form and decompressed before being used. Inter-chiplet transfers are modeled using a modified cycle-accurate HeteroGarnet simulator~\cite{bharadwaj2020kite}, extended to support trace-driven traffic.

\bh{Hardware implementation:}
We implemented the proposed compression/decompression units using SystemVerilog.
It is synthesized at a target operating frequency of 1 \si{\giga\hertz} using the Synopsis Design Compiler with the GF 22 \si{\nano\meter} technology node.




\begin{figure}[b!]
\centering
\vspace{-3mm}
\includegraphics[width=1\linewidth]{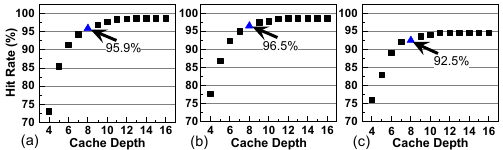}
\vspace{-6mm}
\caption{Local cache hit rate vs. cache depth on WikiText-2 for three models: (a) Jamba, (b) Zamba, (c) Qwen}
\label{fig:hit_rate}
\end{figure}

\subsection{Area-Performance Trade-off Optimization}\label{ssec:trade-off}

\begin{figure}[t]
\centering
\includegraphics[width=0.9\linewidth]{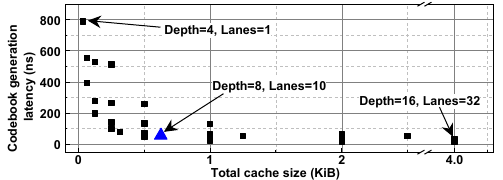}
\vspace{-3mm}
\caption{Codebook generation latency vs. cache size with 512 activations in BF16.}
\vspace{-5mm}
\label{fig:encoding_unit}
\end{figure}

\bh{Compressor trade-offs:} The compressor consists of multi-lane caches, which have two knobs: per-lane cache depth and the number of lanes. 
Figure~\ref{fig:hit_rate} plots the local cache hit rate versus its depth during the encoding phase. 
A depth of 8 entries per lane achieves over 90\% average hit rate for all tested models. Given the diminishing returns of larger sizes, we adopt 8-entry caches in each lane.
Next, we analyze the total cache size to optimize codebook creation latency.  Figure~\ref{fig:encoding_unit} shows this trade-off for the codebook generation latency with 512 activations. While a single lane with depth 4 is area-efficient, it requires 788 \si{\nano\second} to complete at 1 \si{\giga\hertz}. At the other extreme, 32 lanes with depth 16 reduce the latency to 17 \si{\nano\second}, but the total cache size grows to 4 \si{\kibi\byte}. We select a balanced design point with 10 lanes and depth 8, which achieves codebook creation in about 55 \si{\nano\second} with only 0.625 \si{\kibi\byte} of cache.
\vspace{-1mm}


\vspace{-0.5mm}

\begin{figure}[b]
\centering
\vspace{-4mm}
\includegraphics[width=0.9\linewidth]{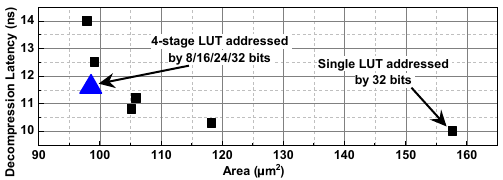}
\vspace{-5mm}
\caption{Average latency to decode 10 exponents versus decoder area. The highlighted point marks our chosen 4-stage LUT decoder with tables addressed by 8/16/24/32-bit prefixes, eight entries per stage.}
\vspace{-1mm}
\label{fig:decoding_unit}
\end{figure}

\bh{Decompressor tradeoffs:}
To analyze the latency–area trade-offs in the decompression unit, 
we implemented it using different multi-stage LUTs configurations. 
As shown in Figure~\ref{fig:decoding_unit}, a 4-stage LUT decoder indexed by 8/16/24/32-bit prefixes achieves an average latency of 11.6 \si{\nano\second} with an area of 98.5 \si{\micro\meter\squared}. For comparison, a single 32-bit LUT table achieves a slightly lower latency of 10 \si{\nano\second} but requires 157.6 \si{\micro\meter\squared} area, making it less area-efficient.
The most optimistic case, required to saturate the $100~Gbps$ link, involves $10$ compressed values (each $10$ bits: $2$-bit codeword + $1$-bit sign + $7$-bit mantissa).
Thus, ten decode lanes we searched based on Figure~\ref{fig:encoding_unit} are sufficient to sustain the full link bandwidth, distributing incoming flits in a round-robin manner. This design balances throughput and cost, delivering near line-rate performance with modest overhead.

\subsection{Communication and End-to-End Execution Time Evaluation}
This section evaluates the BF16 exponent compression ratio (CR) for the Jamba, Zamba, and Qwen models. As shown in Table~\ref{tab:comp_ratio}, LEXI yields the highest CR $\approx3.1\times$, by leveraging per-layer frequency distribution. This significantly surpasses base-delta-immediate (BDI) compression~\cite{pekhimenko2012base} ($2.40\times$), which utilizes micro-local value correlation via $\mathbf{3}$-bit delta encoding. Conversely, the CR of run-length encoding (RLE)~\cite{golomb1966run} $\approx0.64\times$, results in data expansion, confirming that long runs of identical exponent values are infrequent. The primary compression potential thus lies in repeated value frequency, which Huffman coding leverages.

\begin{table}[t]
    \centering
    \caption{Compression Ratio Comparison among different compression methods on different LLM weights} \label{tab:comp_ratio}
    \vspace{-1mm}
    \renewcommand{\arraystretch}{0.6}
    \begin{tabular}{@{} p{2.2cm} c c c c @{}} 
        \toprule
        \textbf{Model/Layer} & \textbf{Base} & \textbf{RLE}~\cite{golomb1966run} & \textbf{BDI}~\cite{pekhimenko2012base} & \textbf{LEXI} \\
        \midrule
        Jamba-mini & $1.00\times$ & $0.62\times$ & $2.43\times$ & $\mathbf{3.14\times}$ \\
        Zamba2-1.2B & $1.00\times$ & $0.65\times$ & $2.36\times$ & $\mathbf{3.07\times}$ \\
        Qwen1.5-1.8B & $1.00\times$ & $0.64\times$ & $2.40\times$ & $\mathbf{3.12\times}$ \\
        \bottomrule
    \end{tabular}
\end{table}

Next, we analyze the impact of LEXI on communication latency and overall execution time. Table $\text{\ref{tab:comm_reduction}}$ reports communication time for uncompressed, weight-only, and LEXI (offline weights $+$ on-the-fly activations/caches). On WikiText-2, uncompressed latency for Jamba is $86.70 \si{\milli\second}$. 
LEXI reduces this latency by $45.4\%$ to $47.35 \si{\milli\second}$, achieving $33.5\%$ and $38.3\%$ reduction for Zamba and Qwen, respectively. On the C4 dataset, LEXI similarly decreases communication by $42.0\%$ for Jamba, $34.0\%$ for Zamba, and $39.2\%$ for Qwen. These results underscore the effectiveness of dynamic Huffman compression during LLM inference.


\vspace{-1mm}
\begin{table}[!t]
    \small
    \centering
    \vspace{-1mm}
    \caption{\textbf{Communication latency (\si{\milli\second})} of Jamba, Zamba, and Qwen models on WikiText-2 and C4 datasets.}
    \label{tab:comm_reduction}
    \vspace{-1mm}
    \begin{tabular}{p{2.0cm} l c c c}
        \toprule
        \textbf{Datasets} & \hspace{-5mm}\textbf{Methods} & \textbf{Jamba} & \textbf{Zamba} & \textbf{Qwen} \\ 
        \midrule
        \multirow{3}{*}{\centering WikiText-2} & \hspace{-5mm}Uncompressed & 86.70 & 8539.52 & 1644.07 \\
        & \hspace{-5mm}Compressed weights  & 80.62 & 8522.16 & 1631.17 \\
        & \hspace{-5mm}\textbf{LEXI} & \textbf{47.35} & \textbf{5682.26} & \textbf{1014.30} \\ 
        \midrule
        \multirow{3}{*}{\centering C4} & \hspace{-5mm}Uncompressed & 294.86 & 12315.53 & 2814.40 \\
        & \hspace{-5mm}Compressed weights & 283.45 & 12262.64 & 2788.61 \\
        & \hspace{-5mm}\textbf{LEXI} & \textbf{171.03} & \textbf{8128.71} & \textbf{1711.34} \\ 
        \bottomrule
    \end{tabular}
\vspace{-2mm}
\end{table}

\begin{figure}[b!]
\centering
\vspace{-2mm}
\includegraphics[width=1\linewidth]{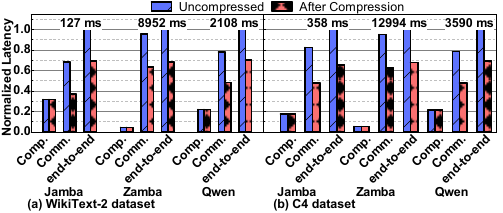}
\vspace{-2mm}
\caption{Normalized average end-to-end latency for Jamba, Zamba, and Qwen models on (a) WikiText-2 with 1K input tokens and (b) C4 with 2K input tokens, each generating 512 output tokens. Annotated numbers indicate the uncompressed end-to-end latency.}
\vspace{-1mm}
\label{fig:e2e}
\end{figure}

Finally, we evaluate the end-to-end execution time on the Simba architecture~\cite{shao2019simba}. 
Since compression does not alter the arithmetic operations, computation latency remains identical in uncompressed and compressed settings. 
Across both datasets, communication latency accounts for 68–95\% of the end-to-end latency in the uncompressed case. 
Figure~\ref{fig:e2e} shows that LEXI 
substantially reduces this bottleneck, delivering 30–35\% reductions in end-to-end inference time across model types and datasets.
Specifically, it lowers the end-to-end latency on the WikiText-2 dataset by 31\%, 32\%, and 30\% for Jamba, Zamba, and Qwen, respectively.
Similarly, the end-to-end latency is reduced by 35\%, 32\%, and 31\% on the C4 dataset.
\vspace{-1mm}

\vspace{-1mm}
\subsection{Hardware Overhead and Power Analysis}
This section evaluates the area and power consumption overheads. 
Table~\ref{tab:area_power} reports post-synthesis results in GF 22 \si{\nano\meter} for the configuration selected in Section~\ref{ssec:trade-off}. 
The compressor employs 10 parallel lanes, each with an 8-entry local frequency cache, alongside a shared global histogram for tree construction. The encoding stage uses 10 parallel LUTs. 
The decompressor is a multi-stage LUT design with four stages, each containing 8 entries and indexed by 8/16/24/32-bit prefixes.

\bh{Compressor:} 
The compression unit includes: (1) 10 local caches (9.85 \si{\micro\meter\squared} each with 2.5 \si{\milli\watt} total power consumption), (2) a single global histogram and codebook generation circuit (13,113 \si{\micro\meter\squared}, 5.23 \si{\milli\watt}), and (3) 10 encoding LUTs (each 79.87 \si{\micro\meter\squared}, with 17.4 \si{\milli\watt} total power consumption). 

\bh{Decompressor:} Multi-stage decoding LUTs occupy 98.5 \si{\micro\meter\squared} area each, for a total 985 \si{\micro\meter\squared} area and 20.3 \si{\milli\watt} power consumption.

\bh{Overall area and comparisons:}
All the circuits within LEXI occupy 14,995.2 \si{\micro\meter\squared} and consume 45.43 \si{\milli\watt} total power. For a fair comparison with the 16 \si{\nano\meter} Simba system, we scale our results using~\cite{stillmaker2017scaling}, resulting in a total LEXI area of 5,452.8 \si{\micro\meter\squared}. Given that each Simba chiplet has 6 \si{\milli\meter\squared} area, the LEXI overhead corresponds to just 0.09\% of chiplet area, demonstrating that the compression/decompression logic introduces negligible hardware overhead.

\begin{table}[t] 
    \small
    \centering
    \caption{Area and power breakdown of compression and decompression units in GF 22 \si{\nano\meter} tech node.}
    \label{tab:area_power}
    \vspace{-3mm}
    \setlength{\tabcolsep}{5pt} 
    
    \begin{tabular}{l c c c c} 
        \toprule
        \textbf{Metric} & \multicolumn{3}{c}{\textbf{Comp. Units}} & \textbf{Dec. Unit} \\ 
        \cmidrule(lr){2-4} \cmidrule(l){5-5} 
        
        & \textbf{\shortstack{Local \\ Cache}} & \textbf{\shortstack{Global Hist. \\ \& Code Gen.}} & \textbf{\shortstack{Enc. \\ LUT}} & \textbf{\shortstack{Dec. \\ LUT}} \\ 
        \midrule
        
        \textbf{Area (\si{\micro\meter\squared})} & 9.85 & 13113 & 79.87 & 98.5 \\
        \textbf{Power (\si{\milli\watt})} & 0.25 & 5.23 & 1.74 & 2.03 \\
        \textbf{Lanes} & $\times 10$ & $\times 1$ & $\times 10$ & $\times 10$ \\ 
        \midrule
        
        \textbf{Total Area (\si{\micro\meter\squared})} & 98.5 & 13113 & 798.7 & 985 \\
        \textbf{Total Power (\si{\milli\watt})} & 2.5 & 5.23 & 17.4 & 20.3 \\ 
        \bottomrule
    \end{tabular}
    \setlength{\tabcolsep}{6pt} 
\end{table}
\vspace{-2mm}
\section{Conclusions} \label{sec:V}

This paper presented LEXI, a lightweight, entropy-aware exponent compression framework designed to accelerate LLM inference on chiplet-based systems. 
Bit-level profiling of weights, activations, and hybrid caches shows that exponents require fewer than three bits, on average, revealing substantial compression opportunities. 
To enable real-time operation, we designed a Huffman-based encoding/decoding framework with parallel multi-lane local caches for tree construction and efficient encoder–decoder LUTs. The proposed design generates codebooks and compressed flits at line-rate (i.e., without impacting throughput). 
Integrated into the Simba chiplet architecture, LEXI reduces inter-chiplet communication latency by 33–45\% and improves end-to-end inference time by 30–35\% across Jamba, Zamba, and Qwen models on WikiText-2 and C4 datasets. 
Hardware synthesis in GF 22 \si{\nano\meter} shows that the design incurs only 0.09\% area overhead.
Overall, we showed that lossless compression of BF16 exponents is effective and provides a practical and scalable solution to mitigate the memory wall toward more efficient hybrid LLM inference.

\vspace{1mm}
\noindent \textit{Disclosure}: Dr. Ogras serves as a contractor for Samsung Austin Research \& Development Center and Advanced Computing Lab (SARC/ACL). This relationship has been approved under applicable outside activities policies.
\newpage
\newpage
\bibliographystyle{ACM-Reference-Format}
\bibliography{references/ref}
\end{document}